# Statistical relationships between corresponding authorship, international co-authorship and citation impact of national research systems

Felix de Moya-Anegon*, Vicente P. Guerrero-Bote**, Carmen Lopez-Illescas*** and Henk F. Moed****

*SCImago Group, Madrid, Spain. Email: felix.moya@scimago.es

**SCImago Group, Dept. Information and Communication, University of Extremadura, Badajoz, Spain. Email: guerrero@unex.es

***University Complutense of Madrid. Information Science Faculty. Dept. Information and Library Science. SCImago group. Spain. Email: carmlopz@gmail.com

****Corresponding author. Sapienza University of Rome, Italy. Email: henk.moed@uniroma1.it

## Abstract

This paper presents a statistical analysis of the relationship between three science indicators applied in earlier bibliometric studies, namely research leadership based on corresponding authorship, international collaboration using international co-authorship data, and field-normalized citation impact. Indicators at the level of countries are extracted from the SIR database created by SCImago Research Group from publication records indexed for Elsevier's Scopus. The relationship between authorship and citation-based indicators is found to be complex, as it reflects a country's phase of scientific development and the coverage policy of the database. Moreover, one should distinguish a genuine leadership effect from a purely statistical effect due to fractional counting. Further analyses at the level of institutions and qualitative validation studies are recommended.

## 1. Introduction

### 1.1  Corresponding authorship and its interpretation

The key concern to disclose the contributions made to scholarly research publications and to properly assess credit and accountability of their authors, led Rennie, Yank and Emanuel (1997) to propose the research guarantor concept. When papers are co-authored certain contributors take on the role of guarantors of the entire research work.

Moya-Anegón et al. (2013) used Rennie, Yank and Emanuel's notion of research guarantorship as a starting point to develop an indicator of research leadership of scientific-scholarly research groups and their institutions. While Rennie, Yank and Emanuel focused on the role of an individual author of a paper reporting on a particular joint research activity, Moya-Anegon et al. aimed to conceptualize and measure, at a higher level of aggregation, leadership roles of research groups conducting joint research projects with other groups. A fundamental assumption is that the research group is the base unit in research, and that research is more and more the result of the collaboration between groups from different institutions.

It is essential to make clear that according to Moya-Anegon et al. (2013), a paper's research leader or guarantor is not necessarily its corresponding author, but the research group or institution to which the corresponding author belongs. Interpreting the corresponding authorship role of a research group as a manifestation of its leadership in institutional collaboration, they applied their method to the global scientific output indexed in Elsevier's Scopus, and compared the output distribution across countries based on corresponding authorship with the distribution based on "whole" or "fractional" counting. The concept of fractional counting is further discussed in Section 1.2 below.

In the paper by Moya-Anegon et al. (2013), the results showed large variations between whole counts and counts based on corresponding authorship. Variability was not found as large when comparing the impact of excellent papers, i.e., the subset of papers among the most frequently cited articles in a particular research field, since the impact of excellent output itself shows less variability as it is based on a selected segment of the citation distribution. They also found that the normalized citation impact of publications as corresponding author tends to be lower than the impact calculated for the total output. Also, the differences between the two impact indicators were found to depend on the international collaboration rates and degree of scientific development of the collaborating countries. The authors concluded that scientifically developed countries benefit more from scientific leadership than developing countries do, while on the other hand developing countries tend to increase their impact with papers presenting research in which they do *not* assume a leadership role.

A validation of corresponding authorship as indicator of the research leadership was published by Sánchez-Jimenez et al. (2016) who presented a case study in the Library and Information Science field. The authors concluded that leadership relationships are useful to weigh the different contributions to research and publications, and better explain institutional and international collaboration results. They characterized leadership not only in terms of the papers for which institutions gained the corresponding authorship, but also in terms of the citation impact or excellence of their collaboration partners. Their study showed that the more important the collaborating institutions are, the higher the acknowledgement to the leading institution. Nevertheless, research leadership cannot be understood in isolation, but as a crucial part of the internal dynamics of institutional collaboration. Sánchez-Jimenez et al. argued that if corresponding authorship of an institution reflects the acknowledgement of its prominent role by the collaborating teams, this recognition is more important, the higher the prominence of the collaborators.

The approach of allocating credits to corresponding authors, institutions or countries has proven to be useful in numerous studies. In an analysis of author citation's practices by Gazni et al. (2016), in several cases the most important role and main responsibility for a paper was only attributed to the corresponding author. The use of corresponding institutions to identify collaborating patterns and fluxes was proposed by Cova et al. (2015). This strengthened the results of the previous mentioned studies relying on the use of corresponding authors to establish the leadership of institutions. In their proposal for the SCImago Institutions Ranking 2012 data visualization, Manganote et al. (2014) also concluded that the leadership indicator provides further insights into the collaborative environment of higher education institutions. The research leadership indicator based on corresponding authorship is applied in many other research works (Álvarez-Betancourt et al. (2014), Olmeda et al. 2016, Guerrero-Bote et al. 2016, Lillo et al. 2013).

Gumpenberger, Hölbling and Gorraiz (2018) linked the interpretation of indicators of corresponding authorship with the Open Access debate. Referring to Machado et al. (2016), they claim that usually the article processing costs (APC) involved in gold open access publishing are paid by the corresponding author's affiliated institution. Although this point is of great interest, and underlines the need to include access modality as a separate factor in further validating the Moya et all. hypothesis of corresponding authorship as an indicator of research leadership, the current authors believe that this phenomenon would by itself not invalidate this hypothesis, as it seems plausible to assume that the leading institution in a collaborative research project would pay the costs of gold OA publishing.

## 1.2 Corresponding authorship and international co-authorship

Many studies explored the use of international co-authorship as indicator of international scholarly collaboration, and underlined the positive effect it has upon the visibility, impact, quality or productivity of research (Narin, Stevens & Whitlow, 1991; Katz & Hicks, 1997; Van Raan, 1998; Glanzel & Schubert, 2004; Lee & Bozeman, 2005; Sánchez-Jiménez, Guerrero-Bote & Moya-Anegón, 2017). Funding programmes such as those of the European Commission promote international collaboration. The percentage of an institution's internationally co-authored papers is a prominent indicator in University Ranking Systems, such as the Leiden Ranking, U-Multirank and SCImago Country and Institute Ranking.

Currently, the role of scientific collaboration studies in the assessment of scientific performance is increasingly growing. The structure of scientific collaboration cannot be explained without taking into account the different roles of the collaborating institutions. Chinchilla-Rodríguez et al. (2016) also explored the Moya-Anegón et al. (2013) method to study the importance of the leading role in collaboration to attain a sound level of scientific performance in the field of nanoscience and nanotechnology. Their findings showed that countries publishing mainly non-collaborative publications tend to have high percentages of total leadership based on corresponding authorship, but low percentages of internationally co-authored papers. Nevertheless, some countries with the highest international collaboration rates present the lowest leadership. Also, high levels of leadership do not always give high levels of performance as a result, and some countries depend more than others on external collaborators to heighten the impact of their research.

In the same line, results from a study by Arencibia et al. (2016), after applying the research leadership indicator based on corresponding authorship, showed that some companies with very high values of leadership have very low of international collaboration. The authors concluded that being the leader of research may be an advantage for some companies, whereas for others the strategy of renouncing leadership to create strong links with institutions may also have positive benefits on their R&D results.

*1.3 Corresponding authorship and whole versus fractional counting of internationally co-authored papers*

Whole counting assigns an internationally co-authored paper fully to each of the contributing countries, while fractional counting assigns only a portion of such a paper to each contributing country. There are several ways to define these portions. Perhaps the most commonly used method is to to assign a paper with *n* contributing countries to each country for a fraction *1/n*. In their paper on the effect of fractional counting upon field-normalized citation impact measures, Waltman and Van Eck (2016) denote this method as "country-level fractional counting".

The indicator based on corresponding authorship explored in the current paper can be seen as a method of fractional counting of internationally co-authored papers, namely as a method assigning such papers fully to the country of the corresponding author, thus allocating a zero portion to all other contributing countries. Waltman and Van Eck label this approach as "corresponding author counting".

Waltman and Van Eck (2016) warned that when the calculation at the level of countries of field-normalized indicators – such as the Leiden Mean Normalized Citation Score (MNCS) indicator, or the SCImago Normalized Impact indicator used in the current paper – is based on whole (or full) counting, the outcomes tend to suffer from a bias in favour of subject fields with a high degree of international co-authorship. This bias is caused by the fact that internationally co-authored (IC) publications tend to be cited more often than non-IC articles, and that in the case of full counting they are counted *multiple* times, once for each contributing country, while in the calculation of the world average they are counted only *once*. In their view, fractional counting does provide results that are properly field normalized.

Waltman & Van Eck's warnings are to the point, and their recommendations have a sound statistical basis. They rightly underline that the reference value of 1.0 for a field-normalized impact indicator based on full counting does *not* properly reflect a world average. Equally important, they stress that the amount of bias varies across research fields. In Section 4.5 the outcomes obtained in the current paper will be discussed from the viewpoint of fractional versus whole counting.

*1.4 Research questions and structure of the paper*

While the studies on the relationship between research leadership and scholarly collaboration mentioned above did *not* distinguish between international and national scholarly collaboration, and analysed particular research areas, the current paper focuses on the relationship between research leadership and *international* collaboration. Moreover, it analyses the global research system *as a whole*, and presents results for a large subset of countries, containing not only scientifically developed, but also developing ones. In order to properly study the relationship between research leadership and citation impact, the current paper also addresses the relationship between international co-authorship and citation impact. The following research questions are addressed.

i. *The relationship between the degree of corresponding authorship and the degree of international co-authorship.* Do countries with a large degree of international co-authorship tend to publish a larger fraction of papers in which they have the corresponding authorship than countries do with less international co-authorships?

ii. *Are internationally co-authored articles cited on average more often than non-IC articles?* How does the normalized citation impact of a country's internationally co-authored articles compare with the impact of its nationally co-authored papers and with articles involving no institutional collaboration at all?

iii. *The relationship between research corresponding authorship and citation impact.* What is the relationship between a country's degree of corresponding authorship on the one hand, and its citation impact on the other? This relationship is analysed for the set of *all* articles published by a country, and for its *internationally co-authored* papers.

The structure of this paper is as follows. Data collection and methodology are outlined in Section 2, and empirical results are presented in Section 3. Finally, Section 4 draws conclusions and makes suggestions for further research.

## 2. Data and methodology

### 2.1 Scimago Institute Rank (SIR)

All bibliometric indicators presented in this paper were extracted from a database named SCImago Institute Rank (SIR) created by SCImago Research Group from publication records indexed for Elsevier's Scopus. SCImago Research Group (SRG) was founded by Félix de Moya in 2007. The SRG is devoted to the study of the scholarly communication system and the development of tools to analyze, visualize and interpret the data extracted from scientific information databases. The SIR Report is a research evaluation Platform and Ranking Generator to analyse research outputs of universities and research-focused institutions. The SIR is a classification of academic and research-related institutions ranked by a composite indicator that combines three different sets of indicators based on research performance, innovation outputs and societal impact measured by their web visibility.

### 2.2 Research indicators

The current paper focuses on collaboration and on normalized impact. It inguishes three types of collaboration: international, national, and 'no collaboration'. International collaboration involves authors from at least two different countries; national collaboration involves authors from at least two institutions from the same country (these authors are labelled as '*domestic*') and *no* authors from a foreign country. No institutional collaboration indicates papers published by authors from one single institution. If a collaboration involves national *and* international co-authorships, it is categorized both as an international and as a national collaboration.

The Normalized Impact indicator based on citations, denoted with the acronym NI in this paper, is based on the methodology established by the Karolinska Institute, where it is named the "item oriented field normalization score average" (Rehn & Kronman, 2008). The field-normalization is established at the article level, using the subject classifications of journals into 27 disciplines implemented in the Scopus database. The time period taken into account is 2003-2015. For more information the reader is referred to SIR (n.d.).

Table 1. Distribution of number of published documents (2003-2015) across countries and institutions

| Level | N | Type | Mean | P99 | P95 | P90 | P75 | P50 | P25 |
|-------|---|------|------|-----|-----|-----|-----|-----|-----|
| Countries | 240 | Totals | 128,856 | 1,769,245 | 641,058 | 210,712 | 26,668 | 2,575 | 336 |
| | | Average/year | 9,912 | 136,096 | 49,312 | 16,209 | 2,051 | 198 | 26 |
| Institutions | 34,414 | Totals | 1,371 | 27,247 | 5,419 | 2,180 | 394 | 39 | 4 |
| | | Average/year | 105 | 2,096 | 417 | 168 | 30 | 3 | 0.3 |

In the current paper bibliometric indicators for 240 countries and 33,000 institutions and for the time period 2003-2015 were extracted from SIR. The following *document types* are included: articles, reviews, notes and short surveys. Table 1 presents the basic statistics on the publication output of the countries and institutions for which indicators are available. The results presented in Section 3 relate to the sub-set countries in the upper half of the distribution of published documents across countries. As can be deduced from Table 1, this sub-set contains 120 countries, each with at least 2,575 publications during 2003-2015, that is, at least an average of about 200 publications per year. Countries are labelled with their 3-character ISO codes. For the corresponding full names the reader is referred to the ISO website (ISO, n.d.).

## 3. Results

### 3.1 The relationship between the degrees of corresponding authorship and international co-authorship

In the set of the 120 largest countries in terms of total number of articles published, a strong negative linear correlation was found of -0.97 (Pearson's R, p<0.001) between a country's percentage of internationally co-authored (IC) articles and the percentage of its articles in which the country delivers the corresponding authorship (CA). The observed negative correlation is not a surprising finding. If a country would not publish any internationally co-authored papers at all, collaboration links would be purely national or institutional, and hence the percentage of papers for which it delivers the corresponding author would by statistical necessity be 100.

The tendency of scientifically emerging countries to show low IC rates and high CA rates is enforced if the database producer decides to index also more nationally oriented journals from those countries. In a study of publication counts in Scopus related to the time period 1996-2010, mean annual growth rates above 20 per cent were obtained for Malaysia, Iran, Romania and Pakistan, between 10 and 20 per cent for the three BRIC countries Brazil, India and China, and for Thailand, Egypt, Portugal and Greece (Moed, 2011), and between 9 and 10 per cent for Turkey, Taiwan, Greece and South Africa. It is plausible to assume that this increase is at least partly due to the inclusion of a disproportionally large number of national journals from these countries in the Scopus database.

*Figures 1 and 2* show a scatterplot of these two variables, for the set of top 40 largest countries, and for the full 120-set, respectively. The 40 countries in *Figure 1* show a range of values of the percentage of internationally co-authored papers between 10 and 60 per cent. This set was divided into two subgroups, one with relatively low (Group 1), and one with high values (Group 2) of the two indicators, based on visual inspection. Group I contains a series of countries that are in a consolidation and expansion phase of their scientific development. They rapidly develop their own infrastructure and become less dependent upon the input of foreign institutions. Hence, the percentage of internationally co-authored articles tends to be relatively low, and domestic researchers acquired corresponding authorship of a large fraction of their output. The four BRIC countries (Brazil, Russia, India and China) are in this group, as well as the Asian countries Japan, Korea, Taiwan. Most other countries in this group are included the list presented above of countries showing large annual growth rates in their Scopus publication counts; their position may be partly due to the inclusion of nationally oriented journals in Scopus. But group I is inhomogeneous, not only in terms of geographical location – as it includes countries both from Asia, Europe and Northern America –, but also in terms of a country's level scientific development; for instance, it includes both BRIC countries but also established economies USA and Japan

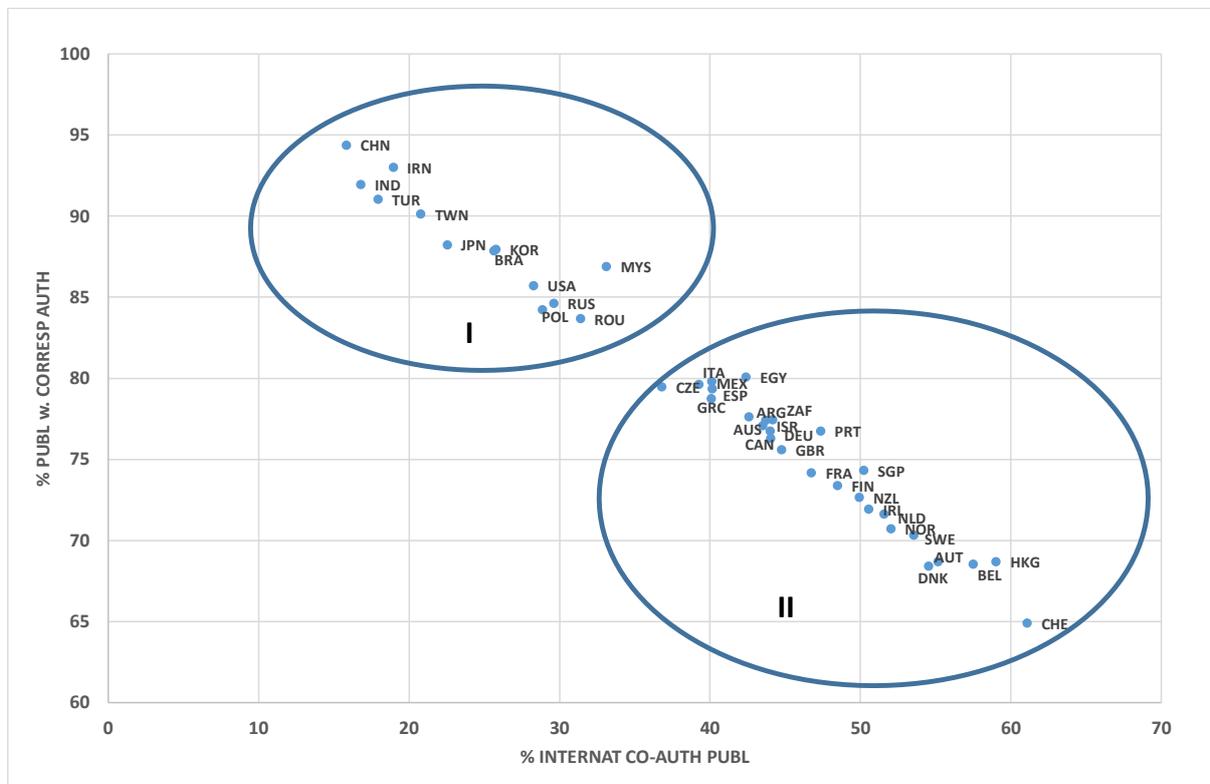

Figure 1. % Papers with corresponding authorship versus % internationally co-authored papers for the 40 largest countries in terms of total number of articles published. Please note that in order to increase the readability of the graph, the vertical axis covers the range 60-100 % only.

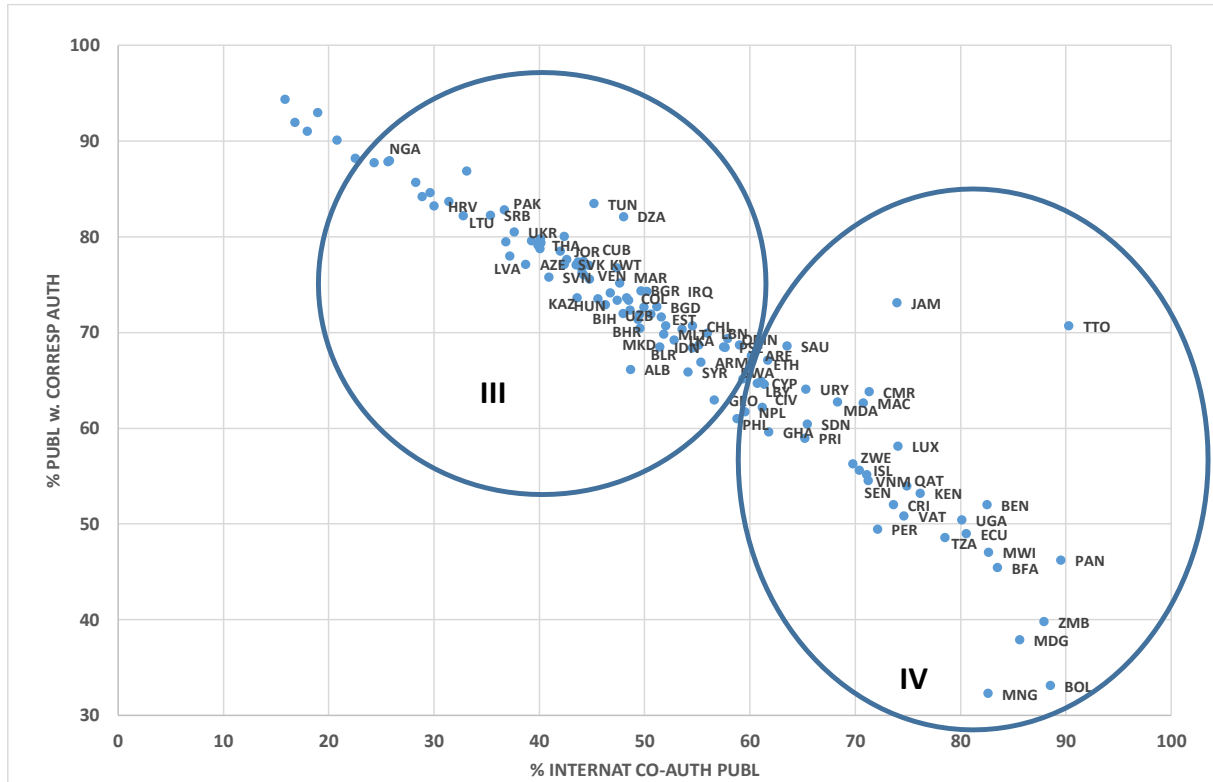

Figure 2. % Papers with corresponding authorship versus % internationally co-authored papers for the 120 largest countries in terms of total number of articles published. Please note that the vertical axis covers the range 30-100 % only. The dots with missing country names are those displayed in Figure 1.

The core of Group II consists of a subset of smaller, scientifically developed European and Asian countries that tend to have a large citation impact and a high degree of international collaboration. Typical examples are Denmark, Belgium and Hong Kong (currently a part of China). But also the larger European countries in term of output, such Great Britain and Germany, as well as Canada and Australia are in this group, be it with a slightly lower international co-authorship and higher corresponding authorship rates.

Figure 2 plots the results for the countries with rank 41-120 in terms of publication counts. Also in this figure two groups are identified based on their degree of internationally co-authored papers, labelled as Group III and Group IV. While Group III seems to be a rather heterogeneous group of countries in their *building-up* or *consolidation and expansion* phase, Group IV contains scientifically mainly developing countries that are in the *building-up phase* of their development. They start participating in international networks, but their role still tends to be secondary, and in relatively few cases they deliver the corresponding author. Figures 1 and 2 do not show any results for 120 smallest countries in terms of publication output. This subset shows a large scatter in which hardly any pattern can be detected on the basis of visual inspection.

### 3.2 Are internationally co-authored (IC) articles on average more cited than non-IC articles?

**Figure 3** shows for the 40 countries with the largest number of published articles the distribution of these articles across the three types of collaboration: international (IC), national (NC), and no (or without) institutional collaboration (WC). **Figure 4** presents the values of the countries' field-normalized citation impact (NI) of each of these types of collaboration, as well as for the aggregate set of all publications, labelled with the acronym 'all'. For *each* country the impact of its IC papers is higher than that of NC articles and non-collaborative articles. This is also true for the wider set of 120 countries. The inequality **Ni.ic>NI.all>NI.nc>NI.wc** holds for most countries. Perhaps the largest deviation occurs for the USA, for which **NI.nc** is much larger than **NI.all**. It must also be noted that the citation impact of *national* collaboration within the USA is by far larger than that of any other country.

The degree of international collaboration is expressed as the percentage of IC papers relative to the total publication output. In the set of 120 countries it is found that the Pearson coefficient of the correlation between the percentage of IC papers and the normalized citation impact of a country's *total publication output* amounts to 0.40 ($p<0.001$). In this sense, more international collaboration is associated with higher citation impact. But, interestingly, Pearson's R for the correlation between the degree of a country's international collaboration and the impact of its *internationally co-authored* papers is almost zero (R=0.04, $p<0.61$).

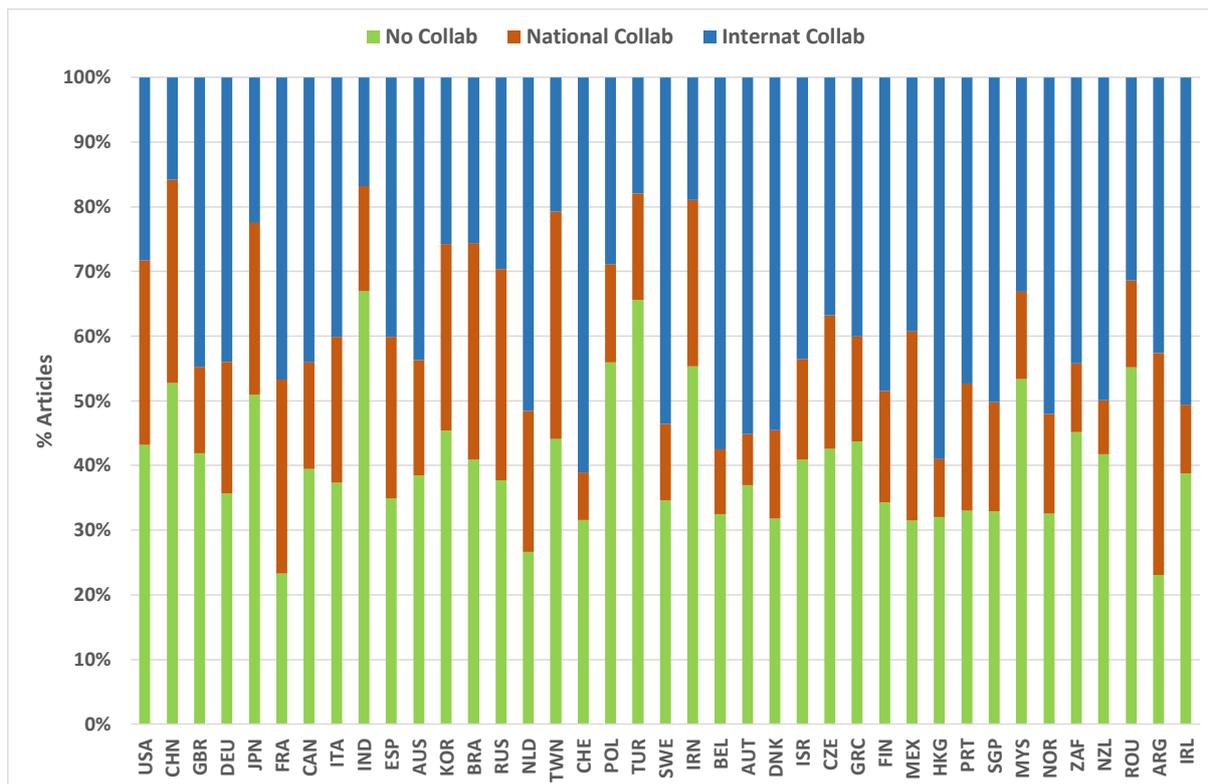

Figure 3: Distribution of articles across type of collaboration for 40 selected countries

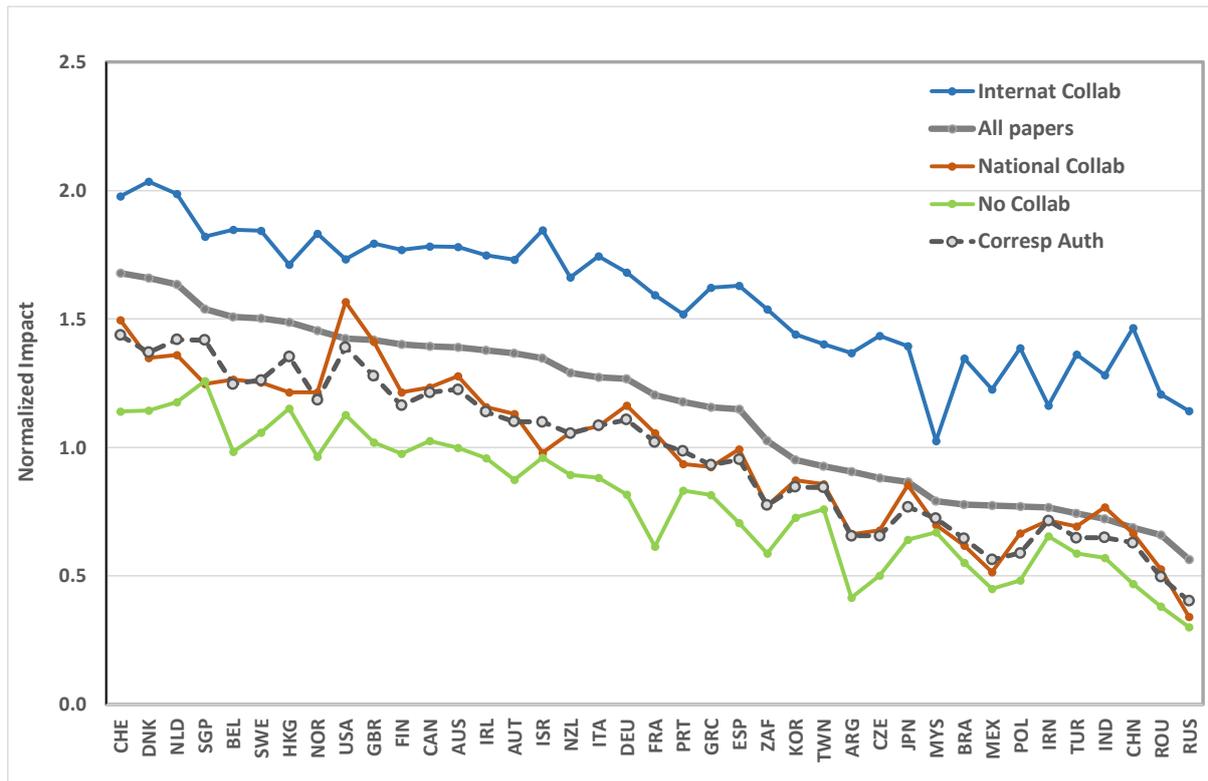

Figure 4. Normalized impact per type of collaboration for 40 selected countries

*3.3 Are a country's papers with corresponding authorship more cited than all its papers?*

**Figure 4** also displays the field-normalized citation impact of the papers with *corresponding authorship* published by each country. The values are similar to those related to the articles resulting from *national* collaboration, except for the USA and India, and much smaller than those of all papers published by a country. **Figure 5** presents a scatter plot comparing the field-normalized impact (denoted as "impact" in Figure 5) of all papers published by a country with that of the articles of which it provides the corresponding author. On average, the values of the latter are 15 per cent *lower* than those on the former.

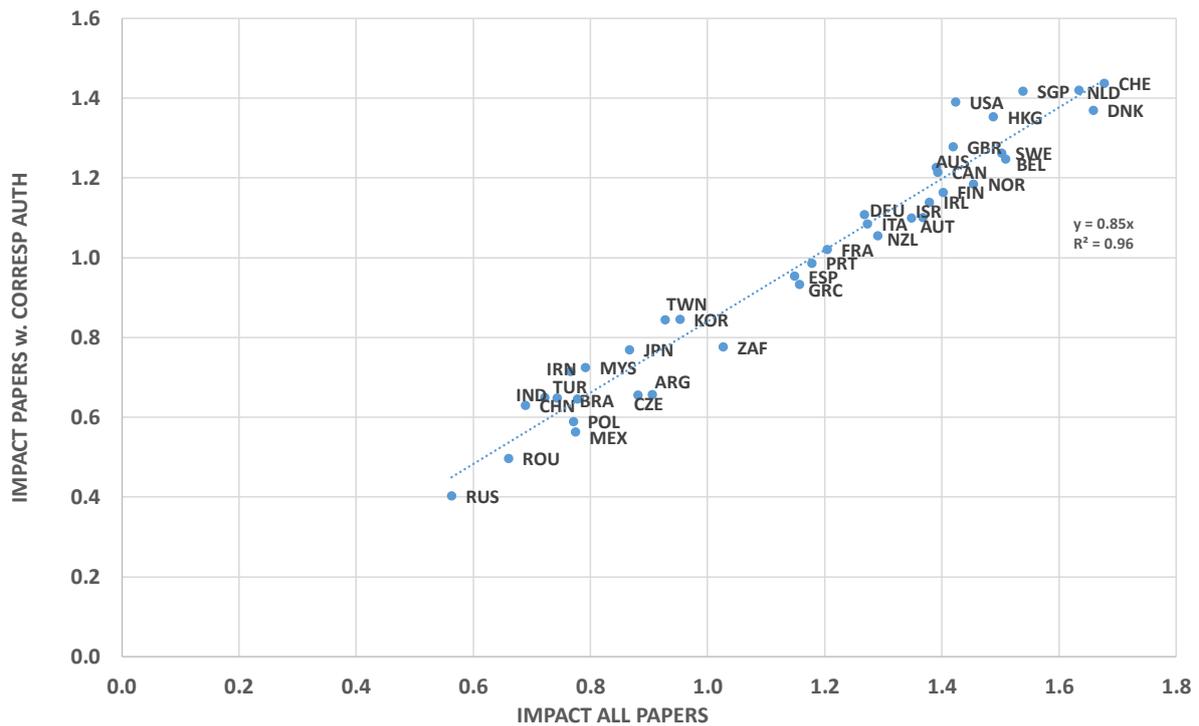

Figure 5. Impact of all papers versus that of papers published as corresponding author (40-country set)

This question is also explicitly addressed in Moya-Anegon et al. (2013). Figure 3 in this article shows for a set of 76 major countries, and publications and citations during 2003-2010, a strong linear correlation (Pearson) of 0.96. Analysing in the current dataset the correlation between normalized impact of all papers, and that of CA papers, Pearson coefficients of 0.98, 0.98 and 0.89 were obtained for the 25, 40 and 120 most productive countries in terms of total number of published articles, respectively.

### 3.4 Are a country's IC papers with corresponding authorship more cited than its other IC papers?

This section focuses on *internationally co-authored* (**IC**) articles. ***Figure 6*** shows for the sub-set of 40 countries the percentage of corresponding authorships among the internationally co-authored articles. For most countries this percentage does not deviate much from 50, but for China, Iran and Malaysia it is substantially larger. As suggested in Section 3.1, this may be at least partly due to the inclusion of a disproportionally large number of nationally oriented journals in Scopus. ***Figure 7*** shows the citation impact of the two subsets of internationally co-authored papers (***NI.ic.ca*** and ***NI.ic.nonca***), and that of all IC publications (***NI.ic***). For all countries except one the following inequality holds: ***NI.ic.ca<NI.ic<NI.ic.nonca***. The only exception is (again) the USA. For this country the impact of IC papers with corresponding authorship fairly exceeds that of the IC papers in which it did not provide the corresponding author.

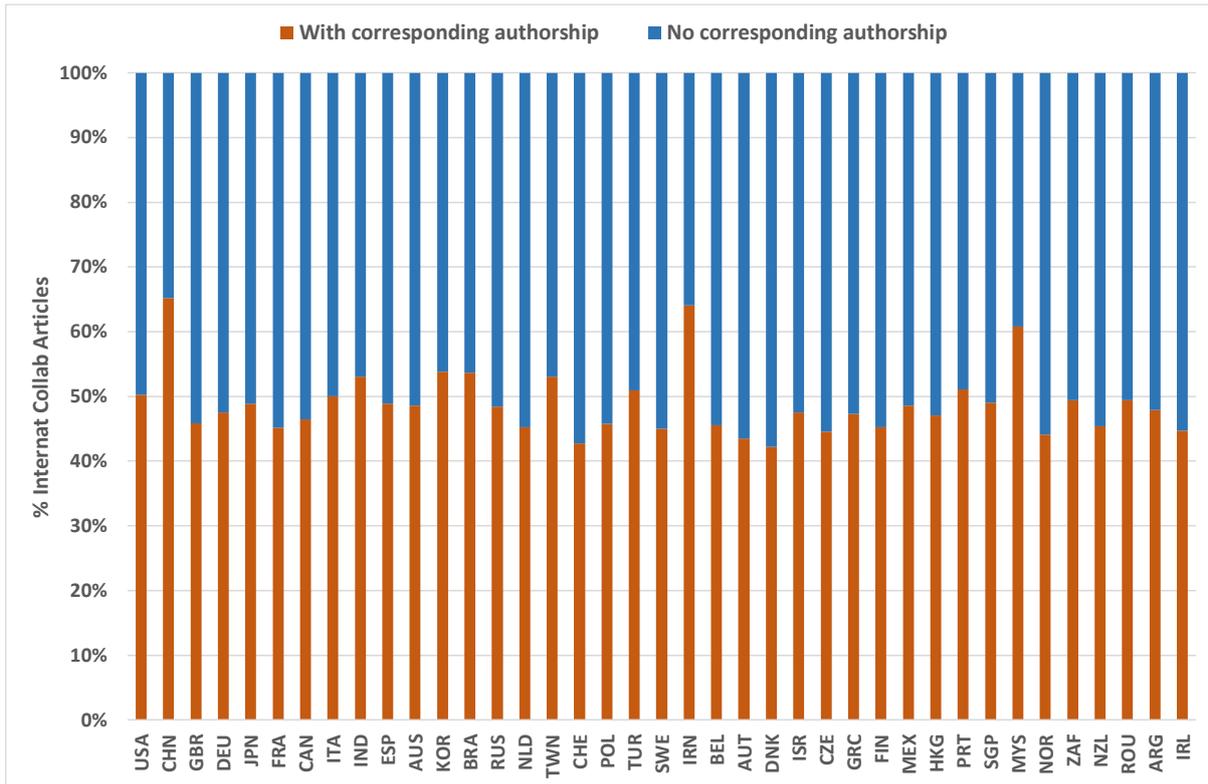

Figure 6. Distribution of internationally co-authored papers across types of authorship for 40 selected countries

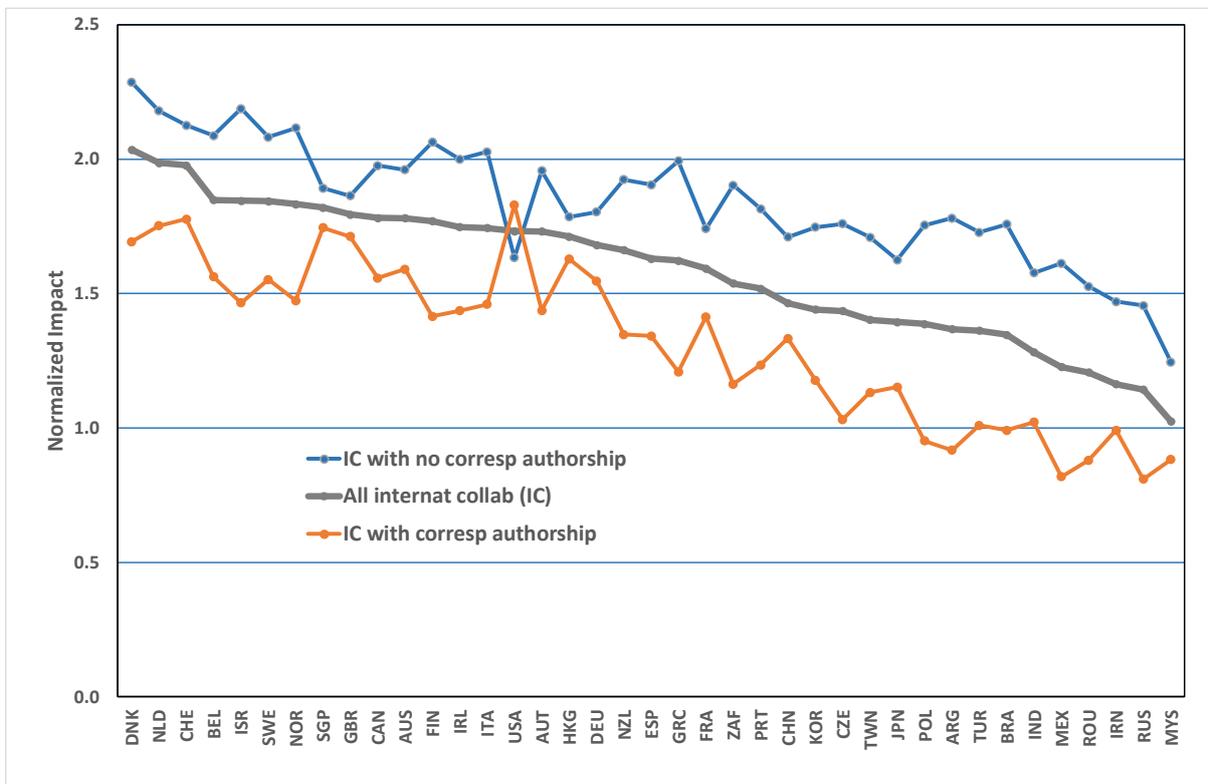

Figure 7. Normalized impact per type of authorship for 40 selected countries

Considering the total set of articles published, the percentage of papers with corresponding authorship show a Pearson R of 0.39 (p<0.001) with the overall normalized citation impact. But for internationally co-authored papers the situation is different. The two variables show a negative correlation of -0.17, not significant at p=0.01.

A scatterplot of the underlying data points is presented in **Figure 8**. It shows that the overwhelming part of the countries has a citation impact of its IC papers above 1.0, often denoted as the 'world average'. This outcome is consistent with the results presented in Figure 4 as regards the citation impact of the 40 largest countries in terms of total publication output. The countries inside the ellipse drawn in Figure 8 have a disproportionally large percentage of corresponding authorships. Among these, the larger countries China, Iran and Malaysia show a strong increase in the annual number of publications indexed in Scopus during the time period considered; many of these are nationally oriented journals, in which institutions from these countries tend to have the corresponding authorship. If one ignores the countries in the ellipse, the overall pattern is similar to that of a random scattering. It is unclear while two Nord-African countries, Tunisia (TUN) and Algeria (DZA) are in this cluster; the same holds for the appearance of the Caribbean countries Trinidad & Tobago (TTO) and Jamaica (JAM).

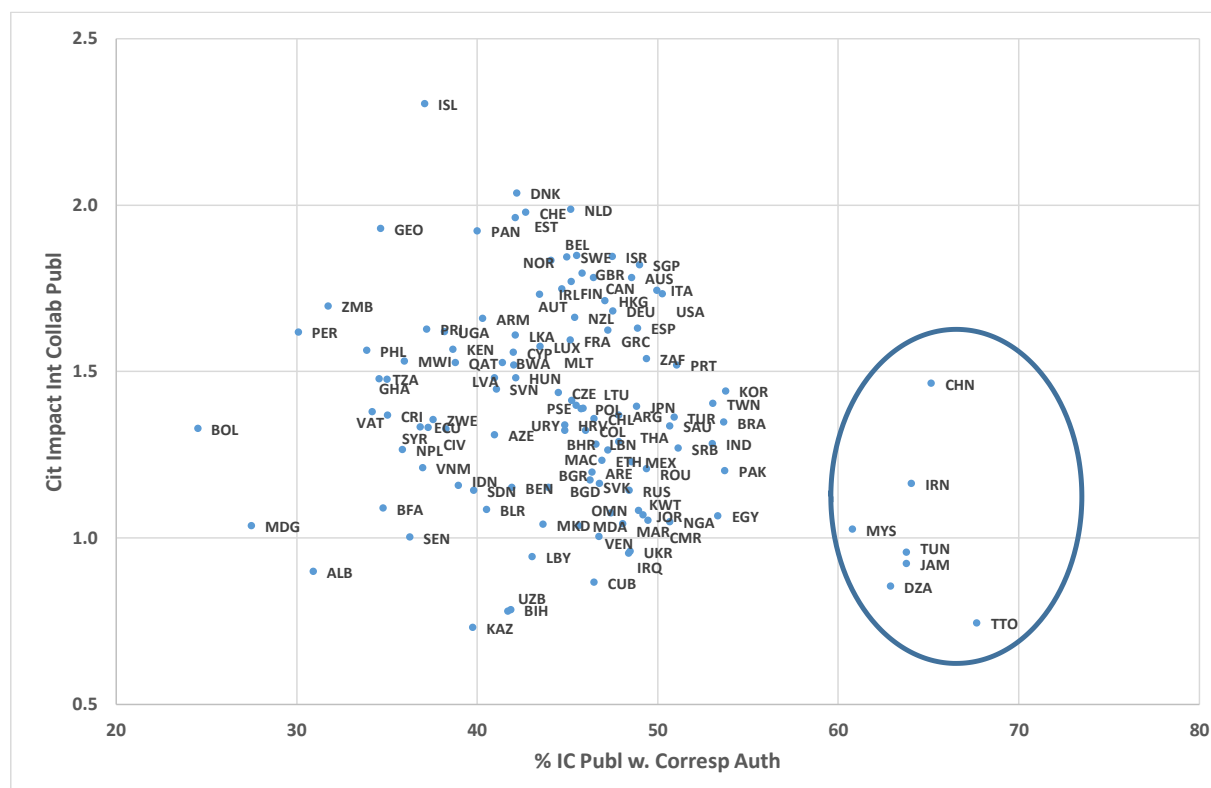

Figure 8. Citation impact of internationally co-authored papers versus the percentage of IC publications with corresponding authorship. Please note that in order to increase the readability of the graph, the vertical axis covers the range 0.5-2.5 and the horizontal axis a range 20-80.

# 4. Discussion and conclusions

## 4.1 The relationship between the degrees of corresponding authorship and international co-authorship

The results obtained in Section 3 illustrate that the degree of international co-authorship and the corresponding authorship rate are to some extent statistically dependent. Countries with a low international co-authorship necessarily show a large corresponding authorship rate. This explains at least partly the strong negative correlation of these two variables observed in the 120-country set.

A recent study by Moed (2016) showed that one can give a meaningful interpretation to a percentage of internationally co-authored papers calculated for a particular country or institution only if one takes into account the *phase of development* of these entities. A large percentage may indicate both an advanced and an early phase of development, and a declining percentage over time may indicate a positive development and strength rather than weakness. The study makes an analytical distinction – not necessarily reflecting a country's actual, chronological development – between a pre-development phase, a building up phase, consolidation & expansion, and an internationalization phase. The observed pattern in Section 3 does reflect to some extent the various phases in this model. There is a group of countries showing high IC rates but relatively low CA rates. These countries are still in the building-up phase of their national research system, and their role in the international collaborations tends to be secondary. The group of countries with low IC rate and high CA rates is more heterogeneous, as it contains the USA, but also Japan, three BRIC countries, and several other expanding nations.

But there is evidence that there is a second factor at stake, namely the extent to which the database (in this case Scopus) covers more nationally oriented journals of countries that are in an expanding or consolidating phase of development. When researchers from these countries participate in an international collaboration, the resulting papers will not often be published in national journals, regardless of whether their role is primary or secondary. As a result, when national journals are indexed, the corresponding authorship rate increases. But one can argue that this factor does not necessarily bias the outcomes if the database expansion itself reflects a country's expansion phase.

The finding that the USA has a relatively low degree of international co-authorship, while its degree of national collaboration is high, is in agreement with Francis Narin's observations (Frame & Carpenter, 1979). Any institution in this large country has many potential collaborators within the USA itself. In fact, they are a set of states. Possibly if the European Union were treated as a single entity, it would show a similar behaviour.

Still focusing on the position of the USA, the low amount of international collaboration and a high normalized impact of the papers with corresponding authorship is in good agreement with previous studies performed at different aggregation levels, such as a study by Benavent-Pérez, Gorraiz, Gumpenberger, et al. (2012) describing the atypical behaviour of Harvard University in comparison with other international top universities.

From a theoretical point of view, indicators based on research leadership are relevant parameters to characterize phases of scientific development, and the indicators explored in the current article are useful to assess a country or institution's state of scientific development, provided that separate indicators are available of the type explored in the current paper, for international and national collaboration.

*4.2  Are internationally co-authored articles on average more cited than non-IC articles?*

The answer to this question is affirmative. The results on the relationship between degree of international collaboration and citation impact confirm the conclusions in earlier studies reporting a positive "effect" of international collaboration upon citation impact (e.g., Narin, Stevens & Whitlow, 1991; Katz & Hicks, 1997; Van Raan, 1998)". It was found that for all countries in the 120-set the impact of their internationally co-authored papers is higher than that of nationally co-authored or non-collaborative articles.

It must be noted that the impact measure used in this analysis is based on whole (or full) counting of IC papers. As argued in Section 1.3, Waltman and Van Eck (2016) rightly underlined that when applying a whole counting scheme, the weighted average impact value across all countries tends to be above one, and in this sense does not properly reflect a world average. Moreover, the size of the deviation of a weighted average from one may differ across research fields. The conclusion obtained in the current paper that IC papers from all countries show an increase in impact relative to their non-IC output is *not* based on the assumption that the 1.0 level does reflect a weighted world average. Differences across subject fields may indeed to some extent distort the picture and, for instance, positively affect the position of countries with a biomedical orientation compared to that of countries specializing in natural sciences, but it is highly unlikely that the overall tendency in the data would change because of this.

*4.3  Are a country's papers with corresponding authorship more cited than all its papers?*

It was found that the normalized citation impact of a country's papers with corresponding authorship is in the set of 40 major countries on average some 15 per cent *lower* than that of all its' papers, and in the 120-set some 25 per cent lower. As argued in Section 1.3, assigning papers to countries on the basis of corresponding authorship can be considered as a special form of fractional counting. Hence, these observed overall differences cannot be interpreted merely in terms of a *research leadership effect*, but are also due to the *statistical effect* of the method of *fractional* assignment embodied in corresponding authorship counting.

The results obtained in the current study do not allow one to separate these two factors. But a rough indication of the relative importance of these two effects can be obtained from data presented by Waltman and Van Eck, who compared in a set of 25 major nations a country's field-normalized citation rates based on whole counting with that based on fractional counting assigning equal fractions to contributing countries, and with the impact derived from corresponding authorship counts (Waltman & Van Eck, 2016, Table A2, data for 25 major countries from the Web of Science for the time period 2010-2013). The latter two variables generated practically identical results: a regression slope of 0.99 and a Pearson correlation coefficient to 0.999. This outcome suggests that, at least at the level of major countries, the difference between impact of all papers and that of papers with corresponding authorship observed in the current article is largely due to a statistical effect.

Waltman and Van Eck also calculated a "full counting bonus", defined as the difference between impact based on whole counts and that derived from fractional count (with equal portions). The values they found in the entire database are similar to the 10-25 per cent obtained for major countries in the current study. This is a *second* indication that the difference between whole counts and corresponding authorship counts are largely due to a statistical effect.

According to Table A2 in the Waltman & Van Eck paper, impact values based on whole counts and fractional counts (based on equal portions) very strongly correlate as well. The current authors obtained a Pearson correlation coefficient of 0.98, equal to the value obtained for the top 25 countries in the SIR-based dataset (the two 25-country sets overlap almost entirely). The Pearson correlation coefficient between these two variables is in the 40-set very high (0.98), and in the 120-set somewhat lower (0.89).

### 4.4 Are a country's IC papers with corresponding authorship more cited than its other IC papers?

The finding that for almost all countries the citation impact of their internationally co-authored papers with corresponding authorship is lower than that of their IC papers in which they do *not* have corresponding authorship illustrates the importance of a country's collaboration partners in building up impactful international research networks.

The finding that in the 120 set a country's share of IC papers with corresponding authorship does not show a significant linear correlation with the citation impact of its IC papers is consistent with Figures 1 and 2. The major part of countries reveals a share of IC articles with corresponding authorship between 40 and 50 per cent. Among the countries in this range, there are both scientifically developed nations participating in impactful international networks, as well as developing countries in their building-up phase, and more involved in regionally oriented relationships that tend to generate as yet a somewhat lower impact. Also, nationally oriented journals from rapidly expanding countries may also publish substantial amounts of collaborative articles with authors from other countries in the region.

### 4.5 Whole versus fractional counting of internationally co-authored papers

When applying a fractional counting scheme by assigning equal fractions of a paper to contributing countries such that the fractions sum up to a value of one, the calculation of field-normalized impact indicators does not merely weigh *articles* on the basis of the number of contributing countries, institutions or authors, but also the number of *citations* to these articles. This means that, for instance, if a paper is produced by researchers from 10 different countries, the number of citations it received is reduced with a factor of 10 as well. In this way, the influence of highly cited internationally co-authored papers upon a final citation impact rate is reduced for *each* of the contributing countries, and, thus, in the impact assessment as such.

Also, giving equal weights to each collaborating country involves a certain arbitrariness. Such a method does not take into account any information on the size of contribution researchers from a country have made to the collaborative effort. The great advantage of the approach based on corresponding authorship analysed in the current paper is that it does take into account this type of information. On the other hand, one should realize that according to this approach the impact of highly cited multi-country collaborations is *fully* attributed to one single country.

In view of the strong linear correlations between the various citation-based indicators for major countries obtained in the current study, one may argue that it does not make much difference which indicator one applies. This may be true when comparing major countries on the basis of their total research output. But if one expands the set of countries analysed, the degree of correlation decreases. Further analyses conducted at the level of *research fields* rather than the total database, and for *institutions* rather than countries, are much more *policy-relevant*, especially if they analyse not only international, but also *national* collaboration patterns as well as the relationship between these two types of collaboration.

*4.6 Limitations and follow-up research*

A number of important issues related to the use of the degree corresponding authorship as a measure of research leadership has not been discussed in this paper. Firstly, the theoretical assumptions underlying this use outlined in Section 1.1 may not be equally valid in all domains of science and scholarship, but adequately capture authoring practices only in 'big' science fields with dense collaboration networks and large research teams. Also, the roles of the first author and last author (if different from the corresponding author) is not given attention, despite the fact these roles may be crucial in the research team. Another open issue is how to deal with multiple affiliations of the corresponding author.

A systematic qualitative *validation* of the leadership indicator based on corresponding authorship would be relevant. The objective of such a study, mainly based on interviews, would be to collect, through interviews or questionnaires with researchers from a series of countries and research fields, more insight into authoring practices, and especially the role of corresponding authorship, in a range of scientific-scholarly disciplines. As indicated in Section 1.1, an important factor to be taken into account is the *access modality* of a paper, but also manifestations of *strategic behaviour*, such as the attempt to increase the probability to have a manuscript accepted in the peer review process by assigning the corresponding authorship to the perceived most prestigious institution regardless of the size of its contribution. Such a study would further enhance the acceptability and hence usefulness of these indicators for practitioners in quantitative science studies and research assessment and the research community at large.

# Acknowledgement

The authors wish to thank two anonymous referees for their valuable comments on an earlier version of this paper.